\documentstyle[12pt,aasms4]{article}
\def\simle{\mathrel{{}^<_\sim}}
\def\simge{\mathrel{{}^>_\sim}}
\begin{document}

\title{Is Deuterium in High Redshift Lyman Limit  
Systems Primordial?}
\author{Karsten Jedamzik$^1$ and George M.
Fuller$^2$}
\affil{$^1$Institute of Geophysics and Planetary Physics, Lawrence  
Livermore National Laboratory, Livermore, CA 94550}
\affil{$^2$Department of Physics, University of California, San  
Diego, La Jolla, CA 92093-0319 }
\authoremail{gail@bethe.ucsd.edu}

\begin{abstract}
Detections of deuterium in high redshift Lyman limit absorption 
systems along the line of sight to  
QSOs promise to reveal the primordial deuterium abundance. 
At present, the deuterium abundances (D/H) derived from the very 
few systems
observed are significantly discordant. Assuming the validity of all the data, if 
this discordance does not reflect intrinsic primordial inhomogeneity, then it must 
arise from processes operating after the primordial nucleosynthesis
epoch. We consider processes which might lead to  
significant deuterium production/destruction, yet allow the cloud to  mimick a 
chemically unevolved system. These processes include, for example, 
anomalous/stochastic chemical evolution  
and D/{$^4$He} photo-destruction. In general, we find it unlikely that these 
processes could have altered significantly (D/H) in Lyman limit clouds. 
We argue that chemical evolution scenarios, unless very finely tuned,
cannot account for significant local deuterium depletion since they 
tend to overproduce $^{12}$C, even when allowance is made for 
possible outflow. Similarly, D/{$^4$He} photo-destruction schemes engineered to 
locally produce or destroy deuterium 
founder on the necessity of requiring an improbably 
large $\gamma$-ray source density.
Future observations of (D/H) in Lyman limit systems
may provide important insight into the initial conditions for the
primordial nucleosynthesis process,
early chemical evolution, and the galaxy formation process. 

\end{abstract}

\keywords{cosmology - quasars -  
chemical evolution -  nucleosynthesis,
abundances}

\section{Introduction}

In this letter we explore issues related to the interpretation of  
recent putative observations of deuterium in seemingly chemically  
unevolved hydrogen clouds along the line of sight to QSOs. These  
observations presently do not provide a consistent value for the  
deuterium abundance, D/H, in high redshift Lyman limit sytems.  
Measurements in several clouds suggest a \lq\lq high\rq\rq\ value,
D/H$\sim 2\times{10}^{-4}$ (Songalia {\it et al.} 1994;  
Carswell {\it et al.} 1994;  
Rugers \& Hogan 1996a; Rugers \& Hogan 1996b;  
Carswell {\it et al.} 1996;  
Wampler {\it et al.} 1996); while determinations in two systems yield a  
\lq\lq low\rq\rq\ value, D/H$\sim 2\times{10}^{-5}$  
(Tytler, Fan, \& Burles 1996; 
Burles \& Tytler 1996a).

It is widely accepted that at least some of these observational  
inferences of D/H reflect the primordial value of this quantity at  
the conclusion of the big bang nucleosynthesis (hereafter; BBN) epoch. This  
belief is founded on the absence of viable alternative  
sites/mechanisms which could produce significant amounts of deuterium without 
overproducing other light elements such as $^6$Li, $^7$Li, and $^3$He 
(Epstein, Lattimer, \& Schramm 1976; Sigl {\it et al.} 1995). 
It is also widely noted that the low metallicities inferred for hydrogen clouds 
at high redshift generally imply only negligible amounts of deuterium depletion 
by stars.

It is important to resolve which (if any) of the various inferred D/H values 
represent the cosmic average primordial abundance 
({\it cf.} Cardall \& Fuller 1996; Hata {\it et al.} 1996). 
Here we use the term \lq\lq average\rq\rq\ since, in principle, 
there could exist intrinsic, primordial, super-horizon scale inhomogeneity at 
the BBN epoch 
({\it e.g.}, isocurvature fluctuations 
{\it cf.} Jedamzik \& Fuller 1995). Such intrinsic inhomogeneity could give 
rise to the apparent discordance in observed D/H, but only if the cosmic average 
of this quantity is D/H$\sim {10}^{-4}$ (Jedamzik \& Fuller 1995). 
A real discordance in D/H is, however, not well established by the data. 
If the apparent discordance {\it is} established by future observations, 
and it does not arise from intrinsic inhomogeneity, then it must result 
from processes operating after the BBN epoch.

It may be that the apparent discordance is simply a result of some subset of 
the data being wrong because, for example, hydrogen \lq\lq interlopers\rq\rq\ 
are mistaken
for isotope-shifted Lyman-$\alpha$ lines (Steigman 1994). 
An erroneous (high) D/H would result if a low column density Lyman-$\alpha$ 
forest line by chance happened to reside at the position in velocity space where the 
deuterium isotope-shifted Lyman-$\alpha$ line is expected. From the observed frequency 
of Lyman-$\alpha$ forest lines in quasar spectra 
(Hu {\it et al.} 1995), one can estimate the {\it a priori} 
probability for any one Lyman-limit system (hereafter; LLS) to have such an interloper.
This probability is given by,
\begin{equation}
P\approx 9\times 10^{-3}\biggl({({\rm D}/{\rm H})_p\over 10^{-4}}\biggr)^{-0.46}
\biggl({N_{\rm HI}\over 3\times 10^{17} {\rm cm^{-2}}}\biggr)^{-0.46}
\biggl({1+z\over 4}\biggr)\biggl({R_{\rm v}\over 10\ {\rm km\ s^{-1}}}\biggr)
\Bigl(1+\xi(\Delta v)\Bigr)\ ,
\end{equation}
and is seen to depend on the primordial $({\rm D}/{\rm H})_p$ ratio,
the column density $N_{\rm HI}$ and redshift $z$ of the LLS, and the observational 
velocity resolution $R_{\rm v}$. The quantity $1+\xi(\Delta v)$ accounts for the
possibility that Lyman-forest clouds may be \lq\lq clustered\rq\rq\ in velocity
space around LLSs. A similar quantity, the clustering of forest clouds around
each other, has been observationally estimated to be approximately $\xi(\Delta v)\sim
1$ for absorber velocity separations $\Delta {\rm v}\simle 100$km s$^{-1}$ 
(Chernomordik 1995; Meiksin \& Bouchet 1995). 
In practice, there is a strong observational bias to claim deuterium detections in
only those clouds which show the smallest Doppler broadening of absorption lines. 
The expected narrow width of the deuterium line, as well as the relative widths of 
the deuterium and hydrogen lines, may then be used to argue against the
interloper possibility on statistical grounds (Rugers \& Hogan
1996a; Burles \& Tytler 1996b).

Even should this issue be resolved, there are a plethora of usually hidden and 
implicit assumptions and decisions which must be made in any assessment of the 
observational data to extract a {\it primordial} D/H. These assumptions revolve 
around issues of chemical evolution and formation histories of LLSs which show 
deuterium. Any such assumptions may be worrisome, given that
even such basic aspects of LLSs as morphology, environment, and their masses are 
poorly understood. LLSs are clouds or sheets of highly ionized gas with temperatures 
around a few times $10^4$ K and with approximate neutral column
densities $N_{\rm HI}\approx 3\times 10^{17}$cm$^{-2}$. It is commonly assumed that 
the bulk of the gas in LLSs is ionized by the diffuse UV background at high redshift. 
Nevertheless, local sources for the ionizing radiation such as young blue stars 
(York {\it et al.} 1990;
Gruenwald \& Viegas 1993) 
or hot galactic halo gas (Viegas \& Fria\c{c}a 1995) have also been 
proposed. It is even difficult to eliminate entirely the possibility that a particular 
LLS is the result of looking through the gas of one, or a few, planetary nebulae.

It is instructive to estimate typical parameters of a LLS such as total baryon mass, 
spatial dimension, and total hydrogen density. Under the assumption of heating/cooling 
equilibrium and/or ionization equilibrium of the cloud with the background ionizing 
radiation, and further assuming spherical geometry for the cloud with a line-of-sight 
passing close to the center of the cloud,
one finds: the total baryon mass,
\begin{equation}
M_b\approx 4\times 10^6 M_{\odot}
\biggl({U\over 10^{-3}}\biggr)^{5.2}
\biggl({J_0\over 10^{-21}{\rm ergs\ cm^{-2}s^{-1}Hz^{-1}}}\biggr)^{-2}
\biggl({N_{\rm HI}\over 3\times 10^{17}{\rm cm^{-2}}}\biggr)^3\ ;
\end{equation}
the radius,
\begin{equation}
R\approx 2\ {\rm kpc}
\biggl({U\over 10^{-3}}\biggr)^{2.07}
\biggl({J_0\over 10^{-21}{\rm ergs\ cm^{-2}s^{-1}Hz^{-1}}}\biggr)^{-1}
\biggl({N_{\rm HI}\over 3\times 10^{17}{\rm cm^{-2}}}\biggr)\ ;
\end{equation}
and the total hydrogen density for the cloud,
\begin{equation}
n_H\approx 5\times 10^{-3}{1\over {\rm cm^3}}
\biggl({U\over 10^{-3}}\biggr)^{-1}
\biggl({J_0\over 10^{-21}{\rm ergs\ cm^{-2}s^{-1}Hz^{-1}}}\biggr)\ .
\end{equation}
In these expressions $U$ is the ionization parameter, i.e. the ratio of the density of 
ionizing photons (with energies $E_{\gamma}>13.6$ eV) to the total hydrogen number 
density, and $J_0$ is the specific intensity of ionizing photons at 
$E_{\gamma}=13.6$ eV. The ionization parameter is inferred from either the relative 
abundances of ionization states of \lq\lq metals\rq\rq\ 
or the inferred temperature of the cloud (Donahue \& Shull 1991). Typical 
uncertainties in $U$ are about one order of magnitude implying a five order of 
magnitude uncertainty in the mass scale of a spherical LLS. Similarly, even under
the assumption that the diffuse UV-background is the source of the ionization of the 
cloud, there is considerable uncertainty in $J_0$, translating into uncertainty in the 
basic cloud parameters. We conclude that not only is it difficult to determine the 
masses of the objects in which D/H ratios are observationally inferred, but it is also 
uncertain how to translate these D/H ratios into a cosmic average. In principle, it is 
difficult to rule out very small masses for LLSs. Such
small clouds could have been subject to significant local deuterium destruction or 
production.

Numerical simulations (Cen {\it et al.} 1994; Katz {\it et al.} 1996) suggest that 
there are two broad classes of
hydrogen absorption systems with hydrogen column densities sufficiently high 
$(\simge 3\times 10^{17}{\rm cm^{-2}})$ to be considered Lyman limit absorbers:
(1) \lq\lq field\rq\rq\ clouds which are distinct and isolated from (proto) galactic 
systems; and (2) the tenuous outer regions of an otherwise massive (proto) galactic 
disk or halo. In the first case of isolated field clouds, the geometries are not well 
determined and they could be compact spherical systems or
extended sheets. One may imagine that the formation and chemical evolution histories 
of these  two classes of Lyman limit absorbers are different.  The question of whether 
different chemical evolution histories in clouds could give rise to inhomogeneity in 
the observed $D/H$ ratios requires resolution. 

Chemical evolution calculations are characterized by specifications of an initial mass 
function (IMF) and a star formation rate. We follow the notation of \markcite{MC96} 
Malaney \& Chaboyer (1996) and take the star formation rate $\Psi$ (in Gyr$^{-1}$) and 
the IMF $\phi (m)$ in $M_{\odot}^{-2}$, so
that $\Psi /\Omega_g$ represents a typical inverse time scale for consumption of 
baryons into stars and $m\phi (m) dm$ is the fraction of mass going into stars within
the stellar mass range $m$ and $m+dm$. Here $\Omega_g$ is the fractional contribution 
of cold gas in damped Lyman-$\alpha$ systems to the critical density and takes values 
of $\Omega_g\sim 0.003$ at redshift $z\approx 3-4$ (Lanzetta, Wolfe, \& Turnshek 1995; 
Storrie-Lombardi {\it et al.} 1995). In this notation the evolution of cold gas with 
time can be written as,
\begin{equation}
{d\Omega_g(t)\over dt}=-\Psi (t) +\int_{m_l(t)}^{m_{up}}\bigl(m-m_r)\Psi 
\bigl(t(z)-\tau (m)\bigr)\phi (m)dm\ ,
\end{equation}
whereas the evolution of the deuterium mass fraction $X_D$ with time is given by,
\begin{equation}
{dX_D(t)\over dt}=-{X_D(t)\over \Omega_g(t)}\int_{m_l(t)}^{m_{up}}\bigl(m-m_r)
\Psi\bigl(t(z)-\tau (m)\bigr)\phi (m)dm\ .
\end{equation}
Here $m_{up}$ represents the mass of the largest stars formed, $m_r$ is the 
remnant mass of a star of mass $m$, and $m_l(t)$ is the lowest stellar mass which 
could have returned its gas to the interstellar medium within the age of the universe 
$t(z)$
(i.e. the lifetime $\tau$ of a star with mass $m_l(t)$ has to satisfy  
$\tau (m_l)=t(z)$).

We may approximate the evolution of the deuterium mass fraction if we assume a 
constant star formation rate (and IMF), neglect remnant masses, and 
approximate $\Omega_g$ and $m_l(t)$ as constant. This yields 
$X_D(t)=X_D(0){\rm exp}(-t/\tau_D)$, with $\tau_D$ the typical time scale for 
deuterium destruction,
\begin{equation}
{1\over\tau_D}={\Psi\over\Omega_g}\int_{m_l\bigl(t(z)\bigr)}^{m_{up}}m\phi (m)dm\ ,
\end{equation}
such that $\Psi /\Omega_g$ is the characteristic time scale for incorporation of 
baryons into stars and the integral is the fraction of stellar material which has 
been returned to the ISM by redshift z.

It has become possible recently to derive constraints on the average star formation 
rate and IMF from observations of damped Lyman-$\alpha$ systems  
(Timmes, Lauroesch, \& Truran 1995; Malaney \& Chaboyer 1996). In order to be 
consistent with the observed decline in $\Omega_g(z)$ with decreasing redshift, 
Malaney \& Chaboyer (1996) argue that typical average star formation 
rates are $\Psi\approx 10^{-2.5}$Gyr$^{-1}$ for $3\simle z\simle 4$. Star formation 
rates in this range would imply a characteristic time scale for incorporation of 
baryons into stars of only $\sim 1$Gyr. Discounting the possibility of outflow, and 
assuming IMF's close to standard (Salpeter), the predicted metal enrichment by 
Malaney \& Chaboyer (1996) is also in rough agreement with the 
observed metallicities in damped systems (Lu, Sargent, \& Barlow 1996). However, 
average deuterium destruction factors ${\rm exp}(-t/\tau_D)$ are predicted to be small 
($\sim 1-5\%$) in the redshift range $3\simle z\simle 4$,
mainly because it is thought that only a small fraction of stellar material (0.1-0.2) 
has been returned to the ISM.

The question arises as to how one could change the IMF and/or star formation rate in 
evolving LLSs in order to \lq\lq achieve\rq\rq\ significant deuterium destruction. 
This may be done locally in stochastic chemical evolution scenarios or globally by 
using non-standard chemical evolution scenarios which incorporate, for example, a 
peaked IMF and/or mass/metal outflows. In an example taken from {\it galactic} 
chemical evolution, it has been shown recently
that destruction of deuterium by a factor of 10 between epochs at high redshift and 
the time of solar system fromation may be possible in models which employ an early 
metal-rich galactic wind (Scully {\it et al.} 1996).

Nevertheless, stringent limits can be placed on the maximum possible deuterium 
destruction in individual LLSs at high redshift by stars with masses below 
$M\simle 40M_{\odot}$, provided the abundances of certain key isotopes
are determined confidently. Stars have to be massive enough so that their 
main-sequence lifetimes are shorter than the age of the universe at redshift 
$z\sim 3-4$. This implies that only stars with masses $M\simge 2M_{\odot}$ could
have contributed to a possible deuterium depletion in the interstellar medium. Note 
that this lower mass cutoff is fairly insensitive to the adopted cosmology, the value 
of the Hubble parameter, and the precise redshift of the LLS. Stars in the mass range 
$2M_{\odot}\simle M\simle 4M_{\odot}$$2M_{\odot}\simle M\simle 4M_{\odot}$, are generally
believed to be significant ${}^{12}$C producers. The $^{12}$C is transported to the 
surface of the star during dredge-ups, when the base of the convective zone reaches 
shells which are highly carbon enriched, and subsequently returned to the ISM in 
planetary nebulae ejecta. The ejecta of AGB stars with 
$2M_{\odot}\simle M\simle 4M_{\odot}$ have typical ${}^{12}$C/H ratios which are 
between 0.1 and 10 
times the solar ratio, depending on stellar mass, metallicity, and the details of the 
dredge-up processes (Iben \& Truran 1978; Renzini \& Violi 1981;
Forestini \& Charbonnel 1996). Most models predict 
${}^{12}$C/H
ratios a few times solar. More massive AGB stars, $4M_{\odot}\simle M\simle
8M_{\odot}$, may in fact be net destroyers of ${}^{12}$C/H (cf. Forestini \& Charbonnel
1996).
The ejecta of massive stars $M\simge 8M_{\odot}$, 
which undergo Type II supernova explosions, are generally expected to be enriched 
in ${}^{12}$C, but also heavier isotopes such as ${}^{28}$Si and ${}^{56}$Fe with 
typical mass fractions of one to a few times the corresponding solar mass
fraction (Woosley \& Weaver 1995).  Here production factors become less certain for 
massive stars  $M\simge 30-40M_{\odot}$, in particular for the heavier isotopes.

The observational determination of carbon and silicon abundances in LLSs 
({\it e.g.}, [C/H]=-2.2 and -3.0 for the two clouds in the system at $z=3.572$ 
determined by Tytler {\it et al.} 1996 from the observations of the carbon ionization 
states CII, CIII, and CIV) can be used to constrain stellar deuterium depletion. 
Adopting moderate carbon production of one times solar over the stellar mass range 
$2M_{\odot}\simle M\simle 4M_{\odot}$ and $8M_{\odot}\simle M\simle 40M_{\odot}$, 
and using $[C/H]=-2$ for the LLS, one can 
infer that not more than $\sim 1\%$ of the gas in the LLS could have been cycled 
through stars in the above given mass range. This implies that deuterium depletion by 
most stars with $M\simle 40M_{\odot}$ cannot
exceed about $1\%$. Note that this constraint can {\it not} be circumvented by 
metal-rich winds (outflow), because the {\it same} stars which deplete deuterium also 
produce ${}^{12}$C abundantly. Moreover, low observed ${}^{12}$C abundances
significantly reduces the possibility that a given LLS results from a line-of-sight 
passing through one or a few deuterium-depleted planetary nebulae.

If one imposes the constraint that significant deuterium depletion by stars must have 
occurred, there are only a few, seemingly highly unlikely, possibilities. 
Chemical evolution could have proceeded via a sharply peaked IMF at $M\approx
6M_{\odot}$. Observational consequences of such a scenario may include the
significant enrichment of the LLS in other isotopes, such as ${}^{14}$N.
As a second possibility, a large 
fraction of material may have been cycled through an early generation of
supermassive stars $M\simge 1000M_{\odot}$ which eject a substantial fraction of their 
initial mass in deuterium-depleted radiation-driven winds 
(Fuller, Woosley, \& Weaver 1986) enriched only in $^4$He. Perhaps direct 
inference of black hole remnants is the only way to establish the viability of such a 
scenario. It may be possible that the carbon abundance in a LLS is underestimated, 
since either the dominant carbon ionization state is CI or carbon is depleted on 
grains. Whereas one can place observational constraints on the CI abundance 
(Burles 1996), the existence of dust in LLSs is not easily observationally constrained.
However, it seems unlikely that significant amounts of dust in LLSs could survive 
evaporation by the ambient ionizing radiation field at high redshift. Lastly, it may 
be that carbon production, and particularly the dredge-up processes in AGB stars, are 
not well understood for low-metallicity stars.

Deuterium may also be produced or destroyed by nuclear photo-disintegration in the
presence of a $\gamma$-ray source: 
${}^4$He($\gamma$,pn)${}^2$H; ${}^4$He($\gamma$,${}^2$H)${}^2$H; or 
${}^2$H($\gamma$,n)p.
For most $\gamma$-ray sources, production of $^2$H dominates over destruction because 
the number density of ${}^4$He targets is much larger than that of ${}^2$H targets. 
In fact, ${}^4$He photo-disintegration has been proposed as an efficient non-BBN 
source for deuterium (Gnedin \& Ostriker 1992),  even though it has been 
subsequently shown that 
this would yield anomalously large ${}^3$He/${}^2$H$\sim 10$ ratios in conflict with 
the presolar abundance ratio ${}^3$He/${}^2$H$\sim 1$ (Sigl {\it et al.} 1995). In any 
case, in the absence of direct
${}^3$He abundance determinations, one may posit that a LLS is enhanced (or depleted) 
in deuterium since it had once been close to a powerful $\gamma$-ray source. Assume, 
for example, the existence of a population of $\gamma$-ray bursters at redshift 
$z_b\simle 1000$ each of which
radiates a flux with spectrum hard enough to produce $\gamma$-ray energies slightly 
above the ${}^4$He($\gamma$,${}^2$H)${}^2$H threshold, $E_{th}\approx 23\,{\rm MeV}$. 
In order for these $\gamma$-ray bursters not to overproduce the 
diffuse x/$\gamma$-ray background at the present epoch, the comoving $\gamma$-ray 
burster density has to be smaller than,
\begin{equation}
N_{\gamma}^c\simle {1\over (10{\rm Mpc})^3}{1\over (1+z_b)}
\biggl({j_{\gamma}(z=0,{E_{th}/(1+z_b)})\over 10^{-5}{\rm
MeV^{-1}cm^{-2}s^{-1}sr^{-1}}}\biggr)
\biggl({E\over 10^{60}{\rm ergs}}\biggr)^{-1}\ ,
\end{equation}
where $j_{\gamma}$ is the specific x/$\gamma$-ray intensity at the present epoch
determined at the energy $E_{th}/(1+z_b)$ and $10^{-5}{\rm
MeV^{-1}cm^{-2}s^{-1}sr^{-1}}$ is the approximate present specific intensity at
$E_{\gamma}\approx 20\,{\rm MeV}$ (Fichtel {\it et al.} 1977).
Here $E$ is the total energy in $\gamma$-rays above threshold in a single burst. An 
adopted approximate comoving distance between $\gamma$-ray bursters of 
$r_c\sim 10\,{\rm Mpc}$ should be compared to the maximum distance by which an 
individual LLS could have been separated from a $\gamma$-ray burst in order to still 
have had significant deuterium production by ${}^4$He photo-disintegration. This 
comoving distance is,
\begin{equation}
r_p^c\simle 10^{-2}{\rm kpc}\ (1+z_b)\biggl({E\over 10^{60}{\rm ergs}}\biggr)^{1\over 2}
\biggl({({}^4{\rm He}/{}^2{\rm H})_p\over 2.8\times 10^3}
\biggr)^{1\over 2}\ ,
\end{equation}
where (${}^4$He/${}^2$H)$_p$ is a primordial number ratio. These distances indicate
that significant deuterium production, as well as destruction, by ${}^4$He/${}^2$H
photo-disintegration should be regarded as an improbable process.

Spatially varying (D/H) ratios at high redshift, if they exist, may have their
origin in the intermediate mass scale primordial inhomogeneity of the 
baryon-to-photon ratio. Jedamzik \& Fuller 1995 pointed out that such primordial
isocurvature fluctuations may yield order unity (D/H) fluctuations on galactic
mass scales ($M\simeq 10^{10}-10^{12}M_{\odot}$) and fluctuations in (D/H) by
a factor $\sim 10$ on the post-recombination Jeans mass scale ($M_J\simeq 10^5-10^6
M_{\odot}$). Nevertheless, such scenarios of BBN can only agree with
observationally inferred primordial abundance constraints if a variety of criteria
are met, such as the efficient collapse of high-density regions, the presence
of a cutoff for isocurvature fluctuations on mass scales $M\simle M_J$ 
(cf. Jedamzik \& Fuller 1995; Gnedin, Ostriker, \& Rees 1995; Kurki-Suonio, Jedamzik,
\& Mathews 1996), and the moderate to significant ${}^7$Li-depletion in low-metallicity
PopII stars. Note that contrary to recent claims (Copi, Olive, \& Schramm 1996) models
which predict intrinsic fluctuations in (D/H) on the LLS-scale are {\it not} 
generally ruled
out by the isotropy of the CMBR. Future observations of (D/H) ratios in different LLSs
may constitute the first test for the presence or absence of baryon-to-photon
fluctuations on intermediate mass scales.

In conclusion, it is difficult to envision a {\it compelling} model for differential 
D/H destruction/production in LLSs that could explain the apparent 
observationally-inferred discordance. The logical leading candidate for such a model 
is anomalous/stochastic chemical evolution involving a finely tuned star formation 
rate or IMF. However, we have argued that most of such models may be ruled out by 
$^{12}$C 
overproduction. In any case, future observations of additional LLSs showing deuterium 
may give insight into the resolution of this problem: either (1) mis- identification 
or -analysis of 
deuterium lines in LLSs; (2) super-horizon scale primordial inhomogeneity at the
BBN epoch; or (3) 
very finely tuned IMF and star formation rates ({\it i.e.} quite different from those 
inferred from galactic chemical evolution considerations) in some LLSs. 
With the advent of the Sloan Digital Sky Survey one may expect a substantial increase
in the number of known bright quasars in the near future. 
It has been estimated that this may yield of the order $\sim 100$ LLSs suitable
for the determination of (D/H) ratios (Hogan 1996).
With the help of this data one may gain important 
new insights into chemical/stellar evolution and the galaxy formation problem.

\acknowledgments
We wish to thank B. Balick, S. Burles, C. Cardall, C. Hogan. J. Prochaska, D. Tytler, 
S. Viegas, and A. Wolfe for useful discussions.  
We also acknowledge the hospitality of the Institute for Nuclear Theory at the
University of Washington where a substantial part of this research has been performed.
This work was supported by NSF Grant PHY-9503384 and  NASA Theory Grant  
NAG5-3062 at UCSD, and under the auspices of the US Department of Energy by the 
Lawrence Livermore National Laboratory under contract number W-7405-ENG-48 and
DoE Nuclear Theory grant SF-ENG-48.

\end{document}